\documentclass[11pt,twoside]{article}

\usepackage[dvips]{graphicx}
\usepackage{asp2006}
\usepackage{epsf}
\usepackage{psfig}
\usepackage{lscape}

\begin{document}

\vspace{1.5cm}

{\flushleft To be published in the Proceedings of the Third Meeting on Hot Subdwarfs and Related Objects, Bamberg 2007}

\title{The UV Spectrum of the Galactic Bulge} 
\author{G.~Busso}
\affil{INAF-Osservatorio di Teramo, Via M.~Maggini s.n.c., 64100 Teramo, Italy}
\author{S.~Moehler}
\affil{European Southern Observatory, Karl-Schwarzschild-Strasse 2, 85748 Garching bei M\"unchen, Germany}

\begin{abstract} %%% Abstract to run on from here.
The UV excess shown by elliptical galaxies in their spectra is
believed to be caused by evolved low-mass stars, in particular sdB
stars. The stellar system most similar to the ellipticals for
age and metallicity, in which it is possible to resolve these stars,
is the bulge of our Galaxy. sdB star candidates were observed in the
color magnitude diagram of a bulge region by Zoccali et al.\ (2003).
The follow-up spectroscopic analysis of these stars confirmed that
most of these stars are bulge sdBs, while some candidates turned out
to be disk sdBs or cool stars.  Both spectroscopic and photometric
data and a spectral library are used to construct the integrated
spectrum of the observed bulge region from the UV to the
optical: the stars in the color magnitude diagram are associated to
the library spectra, on the basis of their evolutionary status and
temperature. The total integrated spectrum is obtained as the sum
of the spectra associated to the color magnitude diagram.
 The comparison of the obtained integrated spectrum with old single stellar
population synthetic spectra calculated by Bruzual \& Charlot~(2003)
agrees with age and metallicity of the bulge found by previous work. The bulge integrated
spectrum shows only a very weak UV excess, but a too strict selection
of the sample of the sdB star candidates in the 
color magnitude diagram and the exclusion of post-Asymptotic Giant Branch stars could
have influenced the result.
\end{abstract}

\section{Introduction}\label{sec:intro}

The UV excess that elliptical galaxies and bulge of spiral galaxies
show in their spectra at $\lambda$ shorter than 2300~\AA~ was one of
the most puzzling discoveries in the last 30 years, since these
stellar systems are believed to be old and metal rich, without
young and massive stars emitting most of their flux at short
wavelength.  It is now widely accepted that this UV emission is caused
by evolved low mass stars, in particular Extreme Horizontal Branch
stars (EHB), called also sdB stars from their spectral classification.
These stars are faint in the optical wavelength range and with the current instrumentation
it is impossible to resolve them in the nearest
galaxies.
The stellar system most similar
to the ellipticals for age and metallicity in which it is possible to
resolve sdB stars is the bulge of our Galaxy. A sample of sdBs star
candidates was observed in the Galactic bulge by Zoccali et
al.~(2003)\nocite{zoccali03}, by means of $V$ and $I$ photometry 
of the region MW05 from the ESO Imaging Survey (EIS\footnote{\tt
http://www.eso.org/science/eis/}, the observations were taken with the Wide Field Imager, WFI@2.2m). 
These stars could be either
highly reddened sdBs or cooler stars affected by lower
reddening. A follow-up spectroscopic analysis of these stars has been
 necessary and observations at the Very Large Telescope (VLT)
telescope were obtained. The data reduction and the comparison of the
obtained spectra with models of hot evolved stars confirmed indeed
that most of these stars are bulge sdBs, while some candidates turned
out to be disk sdBs or cool stars (for more details, see Busso et
al.~2005\nocite{busso05}).  To be sure that the observed bulge region
was not peculiar, other bulge fields were searched for sdB
candidates: EIS photometric data of the bulge fields MW07 and MW08
were reduced and analyzed, finding that sdB star candidates are
present also in these fields.

This work presents the procedure adopted (following the recipe as in
Santos et al.~1995) to construct the integrated spectrum of the bulge
region MW05.

\section{Correction for Reddening and for other Contaminating Stellar
  Populations} 
In order to construct the integrated spectrum of the bulge, it is
necessary to correct for the extinction caused by the interstellar
medium and to take in account that along the line of sight we are
observing not only the bulge but also the Galactic disk. Moreover also
a globular cluster (NGC~6558) is present in the observed field.

To correct for reddening the Schlegel et al.~(1998)\nocite{schlegel98}
dust maps were used.  Their resolution is roughly 5 arcmin, that is
about half the size of a WFI chip:
the reddening was calculated then for the 16 regions resulting by
splitting each of the 8 WFI chips in half, correcting in this
way also for differential reddening.

If two stellar populations have different kinematics, the more precise
method to distinguish them is to compare proper motions of the stars,
which can be measured only by comparing the positions of the same
stars at two different epochs.  In this case though, the observations
were taken in one epoch  only.  To subtract the
contaminating stars of the globular cluster NGC~6558 in the MW05
bulge field, a region centered on the center of the globular cluster,
with a radius equal to the tidal radius of the cluster, was
considered. To validate this approach, also radial star counts on the
image were taken in account: in the external regions, roughly beyond
2000 pixels ($\sim$8 arcmin) from the center, the star counts start to flatten,
indicating that there is no significant contribution from the
globular cluster beyond that range. Only stars in this external
region were then taken into account. 

To decontaminate the bulge color magnitude diagram (CMD) from the
foreground disk population, a statistical approach was adopted, using
synthetic CMDs of the disk in the direction of the observed bulge
field.  The disk simulations (one for each WFI chip) were provided by
S.~Ragaini (priv. comm., PhD Thesis at the University of
Padua, for a detailed description of the simulations, see Vallenari et
al.~2000\nocite{vallenari00}, 2006\nocite{vallenari06}).

\begin{figure}[htbp]
\begin{center}
\includegraphics[width=3.5in,height=3.5in]{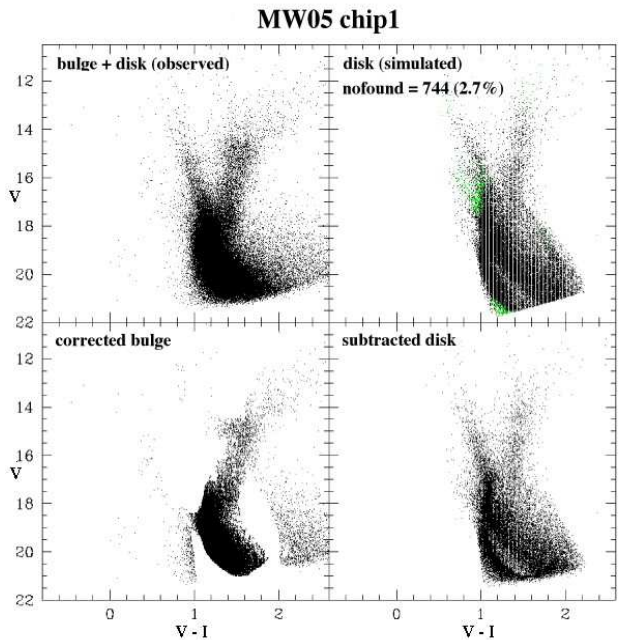}
\caption{Example of the statistical subtraction of the simulated disk
  field from the observed bulge field MW05, in this case for the WFI
  chip \#1. Top left panel: observed bulge field CMD; top right panel:
  simulated disk field CMD, where also the number of simulated disk
  stars not present in the empirical bulge CMD is labeled;
  bottom left panel: bulge CMD after the disk subtraction; bottom
  right panel: subtracted disk CMD}
 \label{fig:disk_sub}
\end{center}
\end{figure}

For each disk star in the disk CMD (Fig.~\ref{fig:disk_sub}, top right
panel) the closest star in the bulge CMD was picked up and
subtracted. The ``photometric'' distance between two stars in the CMD
was defined as:
\begin{eqnarray*}d=\sqrt{[7\times\Delta(V\!-\!I)]^2 + \Delta I^2}. \end{eqnarray*}
and the bulge star with the smallest distance from the disk star was
subtracted. The resulting, cleaned CMD of the bulge is shown in the
left bottom panel of Fig.~\ref{fig:disk_sub}, while the CMD of the
stars statistically removed from the bulge CMD is shown in the right
bottom panel.

%%%%%%%%%% FIG 1 %%%%%%%%%%%%%%%%%%%
%%%%%%%%%%%%%%%%%%%%%%%%%%%%%%%%

\begin{figure}[!tbp]
\begin{center}
\includegraphics[width=3in]{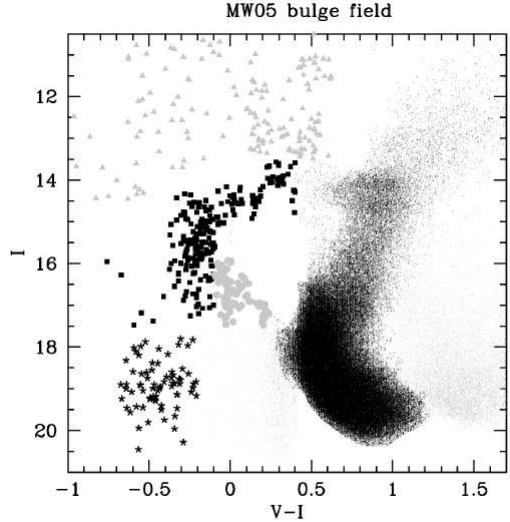} 
\caption{Final color magnitude diagram for the bulge field MW05. See text for
  the stars shown with different symbols. Only stars in black were used to construct the integrated spectrum}
\label{fig:mw05final}
\end{center}
\end{figure}

The remaining bulge stars were combined to obtain the final CMD for
the bulge field MW05 shown in Fig.~\ref{fig:mw05final}. In this figure
the typical bulge sequences, as the Main Sequence (MS), the
red giant branch (RGB), the red horizontal branch (HB) clump (at
$V-I\sim$0.6 and $I\sim$14) are evident.  The grey 
dots are stars that remain after the disk subtraction, because the
simulations do not manage to reproduce well the photometric errors.
The filled grey circles could be Blue Stragglers but previous studies as
Kuijken \& Rich 2001 ruled out their presence in the bulge, therefore they are assumed to be caused by a not precise subtraction.  The
stars marked as filled squares seem to form an extended HB sequence.  Blue HB
stars have been previously observed in the bulge by Peterson et
al.~(2001)\nocite{peterson01} and moreover an extended HB is not typical
for the disk population.  The black triangles are probably a combination of
real post-HB stars and remaining disk blue MS stars.  Since there is
no way to disentangle the two populations in this case and since
post-HB stars should play only a marginal role in the UV excess (Brown
et al.~1997\nocite{brown97}), they were left out from the construction
of the integrated spectrum.  The  stars marked with asterisks represent the sdB star
candidates and they are a mixture of real sdB stars and cooler stars
with lower reddening.  For the construction of the integrated spectrum
therefore only the black (MS, RGB, Red HB, Blue HB and Extended HB) stars were considered.

%%%%%%%%%% FIG 2 %%%%%%%%%%%%%%%%%%%
%%%%%%%%%% %%%%%%%%%%%%%%%%%%%%%%

\section{The Bulge integrated spectrum}

To finally construct the integrated spectrum of the Galactic
  bulge from the UV to the optical the method of Santos et al.~(1995) was
adopted, using both spectroscopic and photometric data and spectral
libraries (Pickles 1998, Lejeune et al.~1997) and the Bamberg archive of
optical and UV-spectra of hot subdwarfs
(Heber, priv.comm.) to extend the sdB spectra to the UV.

To construct the sdB spectrum, first of all it was necessary to decide
which of the sdB star candidates in the CMD 
to take into account. Only stars with a good photometry were selected
(error in the color magnitude diagram  smaller than 0.1), obtaining 112 sdB star
candidates.  The disk simulation, while taking into account the
presence of Horizontal Branch stars, does not include sdB stars which
therefore remain in the corrected disk CMD.  However the analysis of the spectra (see 
Sec.~\ref{sec:intro}) showed that only $\sim$80\% of the
candidates are really sdB stars and not all these stars have turned
out to belong to the Galactic bulge. The percentages found from the
spectroscopic analysis were applied to the total number of sdB
stars candidates, since there is no reason to assume that peculiar
stars were picked up during the selection of the spectroscopic
targets: of 116 candidates therefore, only 55 \% have been considered to be true 
sdB stars of the bulge (Busso et al.\ 2005), the remaining
ones being cool stars or hot stars belonging to the disk or
unclear membership, thus narrowing the sdBs sample to 62 candidates.

%%%%%%%%%% FIG 3 %%%%%%%%%%%%%%%%%%% 
\begin{figure}[htbp]
\begin{center}
\includegraphics[width=3.5in,height=3.5in]{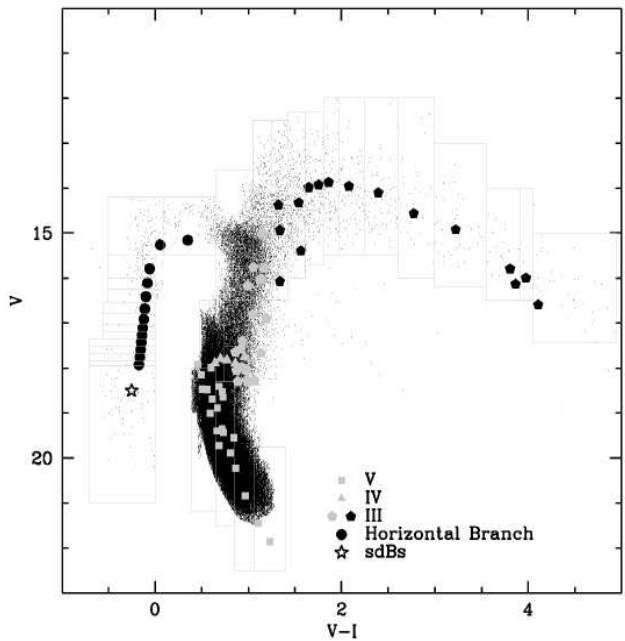}
\caption{Association between library (in this case Pickles 1998) spectra and CMD: the spectra library
  points are marked with different symbols (grey filled squares: dwarfs(V); grey filled triangles:
  subgiant(IV)); grey and black filled pentagons: giants (III); black filled circles : horizontal branch stars; empty star (sdB stars). For each kind of stars, each point has different
  temperature. Every CMD star has been associated to the library
  spectrum corresponding to the box where the star belongs.}
 \label{fig:box_on_cmd}
\end{center}
\end{figure}
%%%%%%%%%% %%%%%%%%%%%%%%%%%%%%%%

In order to associate the appropriate spectrum to the other CMD stars,
the Pickles (1998) spectral library was plotted on the cleaned and
reddening corrected CMD of the bulge region, as shown in
Fig.~\ref{fig:box_on_cmd}.  In this figure, the library points are
marked with different symbols, depending on the luminosity of class
(filled squares: dwarfs~(V); filled triangles: subgiant~(IV)); filled pentagons: giants~(III)) while HB stars are marked with filled circles and sdB stars with an empty star. Each
different point corresponds to a library spectrum with a different
temperature. The boxes on the CMD were chosen so that every box
contains at least one library point. For each box the
following parameters were calculated: the mean absolute
magnitude $<M_{V}>_j$
\footnote{assuming an average distance of 8.15 Kpc, calculated as the most probable distance.}
of the stars in the $j$-th box; the corresponding 
library spectrum ``magnitude'' $m_{V_j}$ obtained from the
convolution with the $V$ filter; the weighting factor $C_j$ of the stars
inside a certain box contributing to the total spectrum ($n_j$ is the
number of stars inside the $j^{th}$ box):
\begin{center}
$C_j = n_j 10^{-0.4(<M_V>_j - m_{Vj})}$
\end{center}
Finally, the total integrated spectrum of this bulge region was calculated as sum of 
all spectra ($f_j$) associated to the CMD boxes, taking in account their weights $C_j$:
\begin{center}
$ \mathcal{F}_{TOT} = \Sigma^N_{n=1} C_{j} f_{j}$
\end{center}

\begin{figure}[htbp]
\begin{center}
\includegraphics[width=80mm,height=59mm]{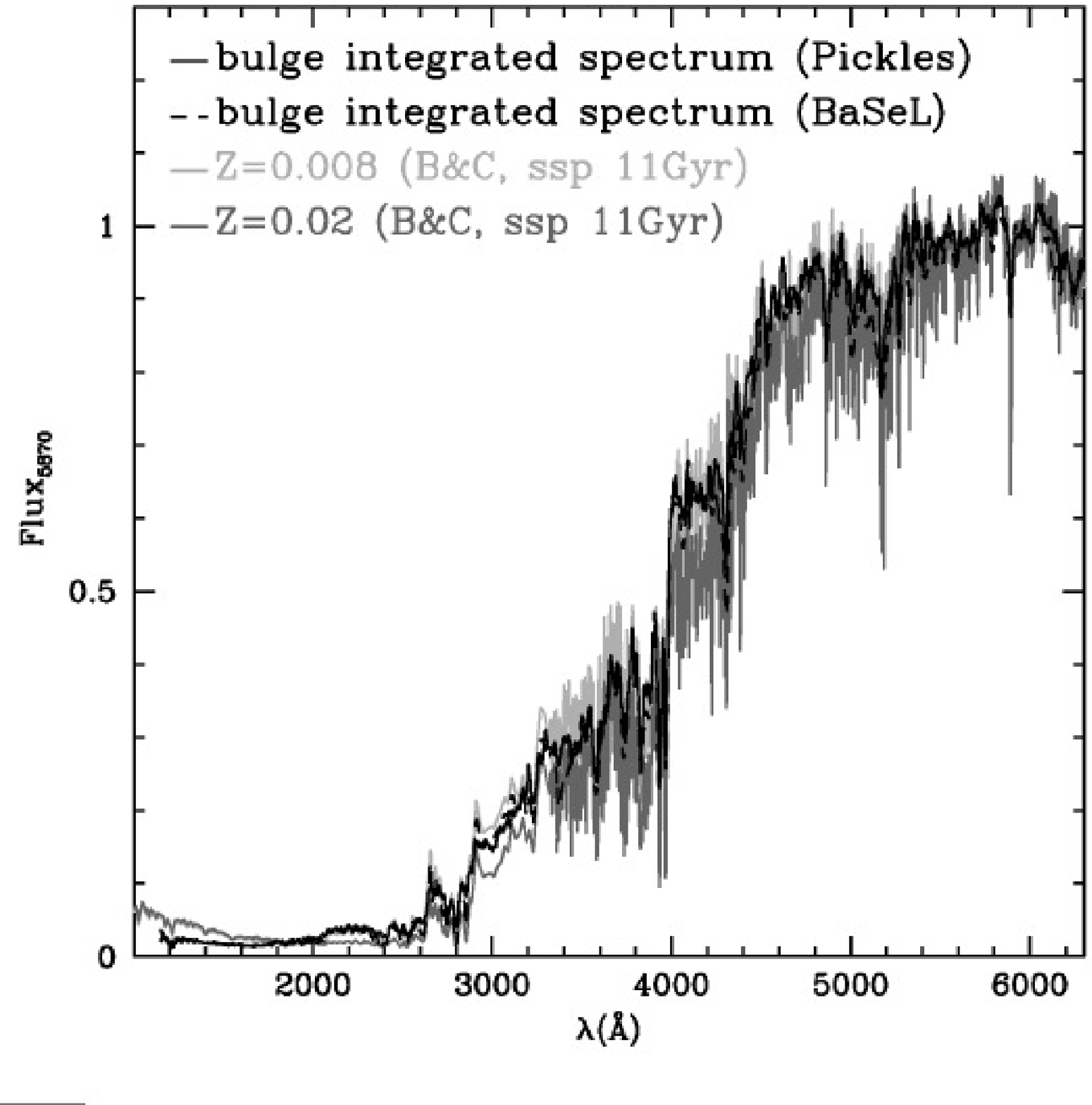}
\includegraphics[width=80mm,height=59mm]{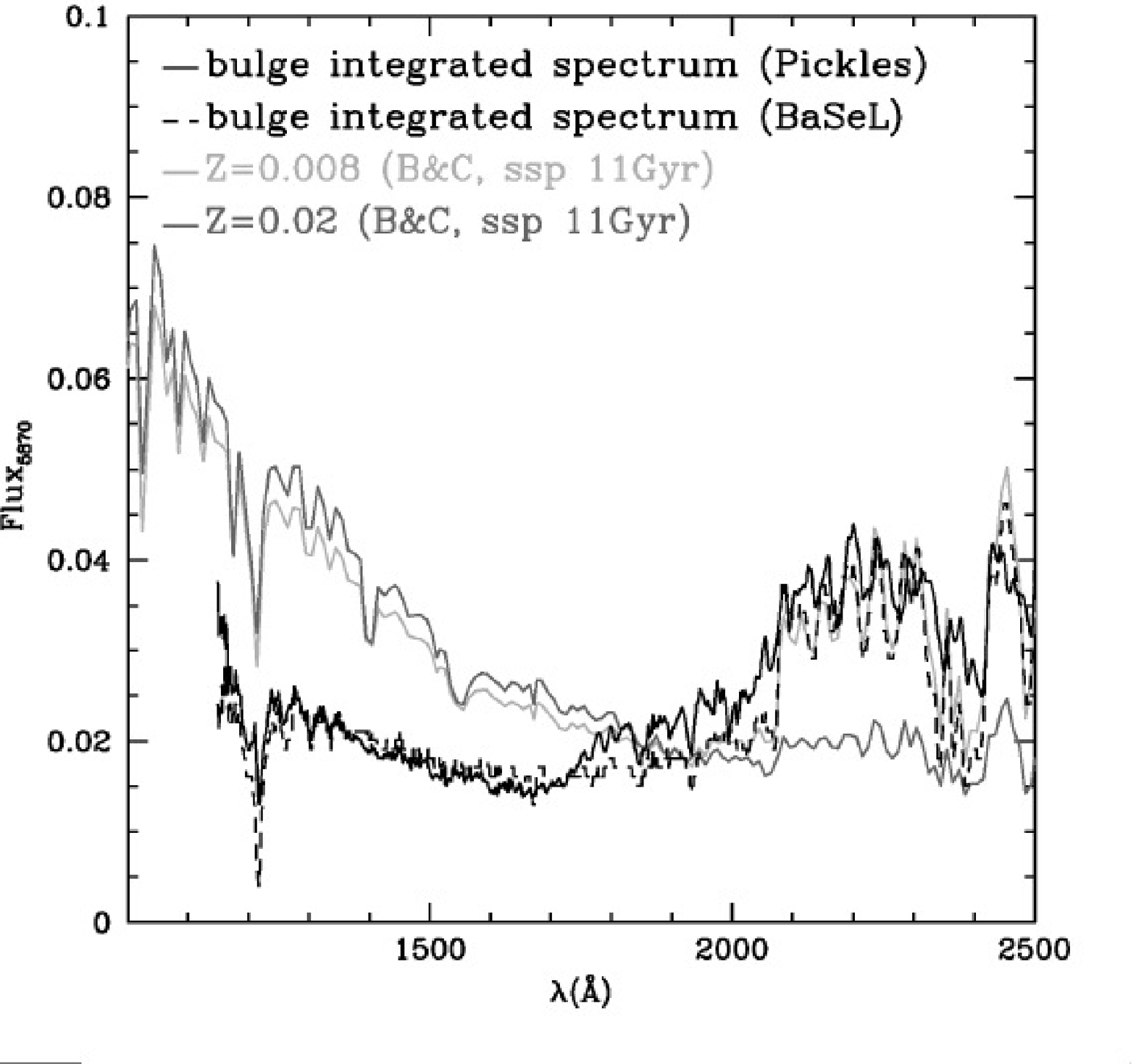}
\caption{Comparison between integrated spectra and single stellar
population SED. Top panel: The integrated bulge spectrum obtained with
the Pickles 1998 and BaSeL (Lejeune et al.~1997) library are shown with a solid and dashed black line respectively. The Bruzual \& Charlot~(2003) spectra
for a single stellar population 11~Gyr old and with metallicity z=0.008
and 0.02 are shown in grey (see label).  To allow the
comparison the spectra were scaled at $\lambda=$5870~\AA.  Bottom
panel: as top panel, but in the UV range.}
 \label{fig:bulge_spek}
\end{center}
\end{figure}

The same procedure was applied also using the BaSeL (Lejeune et al.\
1997) syntethic spectral library, where the spectra at solar
metallicity, with temperature and luminosity class corresponding to
those of the Pickles (1998) stars were selected. The result is shown
in Fig.~\ref{fig:bulge_spek}, where also the comparison with the
integrated spectra (from Bruzual \& Charlot~2003) of two simple
stellar populations (SSPs) with an age of 11 Gyr and metallicity Z=0.008 and
Z=0.02 is shown.

The two integrated spectra obtained using the Pickles and BaSeL
library are very similar, both in the optical and UV range (see
bottom panel).  Neither the SSP
at Z=0.008 ([Fe/H]$\sim -$0.5) nor the one at Z=0.02
([Fe/H]$\sim-$0.09) agree perfectly with the integrated spectra,
particularly in the region between 3300 and 4000~\AA, where the two
synthetic spectra bracket the integrated one. Probably a population
with an intermediate metallicity would fit better. It is necessary to
keep in mind though that the bulge is not a single stellar population
with only one value for the metallicity, but it has a metallicity
distribution from metal-poor (down to [Fe/H]$\sim -$2) to metal-rich (up
to [Fe/H]$\sim$0.5), with a peak at [Fe/H]$\sim -$0.2 (Zoccali et
al.~2003, Rich \& Origlia~2005),
intermediate between the metallicity of the two single stellar
populations adopted.  The expected age for the bulge is 10-13 Gyr
while the adopted single stellar populations are 11 Gyr old. We note
in passing that the fact that the spectrum characteristic of a single
stellar population, hence formed in an "instantaneous" burst, is in
agreement with the recent work of Zoccali et al.~(2006), that, by
means of $\alpha$-element analysis, found that probably the bulge
formed very quickly.

Integrated and synthetic spectra do not match in the UV instead,
with the synthetic spectra showing a larger UV flux. This could be
explained with two reasons. Firstly, in their stellar population
synthesis, Bruzual \& Charlot~(2003) take into account also post-HB
and post-AGB, which have a strong flux in the UV range. These stars
were not accounted for here since it was not possible to
disentangle them from the disk population.  Secondly, probably the
selection of sdBs stars was too strict.
On the base of the
spectroscopic analysis, were selected, as bulge sdBs, on the color magnitude diagram only 55\% of the candidates to construct the
integrated spectrum. Taking into account also the stars with unknown
membership the percentage would raise to 65\%, increasing the UV flux
by about 20\%, which is insufficient to achieve agreement with
the SSP predictions.

Thus the Galactic bulge probably shows only a very weak UV excess,
in contrast to what is observed in the bulge of M31. This is
consistent with UV observations of the closest extragalactic
systems, which show that this UV excess can vary strongly from object
to object (Rich et al.~2005). In addition, the same observations indicate 
also that the UV excess shows no correlation with
physical parameters as metallicity or velocity dispersion. A
correlation with age appears more probable but it would be visible
only at high redshift (Brown et al.~2003).

%%%%%%%%%% FIG 4 %%%%%%%%%%%%%%%%%%%
%%%%%%%%%%%%%%%%%%%%%%%%%%%%%%%%

To verify the procedure, we constructed the integrated spectrum also
for the Galactic Globular Clusters NGC~6388 and NGC~6441, which show
an unexpected population of sdB stars. The procedure was the
same as for the bulge fields, with the same choice of not including
the post-HB and post-AGB stars. The integrated spectra were compared
with the observed integrated spectrum of the two clusters in the
optical (Schiavon et al.~2005) and in the UV (Rich et al.~1993). Also
in these cases, while in the optical the calculated and observed
integrated spectra match, in the UV there seems to be no
agreement. This could be a hint that sdB stars are not the
dominant cause for the UV excess in these systems, while a more
important role could be played by post-HB and post-AGB stars.

In conclusion, the bulge of our Galaxy probably shows only a weak UV
excess. We note that this result is neither in contradiction with
nearby observations, as in the case of M31, nor with observations of
more distant galaxies. In fact the UV excess can vary considerably
from object to object and shows no dependence on physical parameters
like metallicity and velocity dispersion, as explained by Rich et
al.\ (2005).

\acknowledgments %%%text of acknowledgments runs on after this command.
We warmly thanks Manuela Zoccali for providing the photometric data, Uli Heber for the Bamberg spectroscopic archive of sdBs and Sukyoung Yi for the theoretical tracks. 
GB gratefully acknowledges support from the Deutsche Forschungsgemeinschaft through grant Mo 602/8.


\begin{thebibliography}{}

\bibitem[]{brown97} Brown, T.~M., Ferguson, H.~C., Davidsen, A.~F., \& Dorman, B., 1997, ApJ, 482, 685
\bibitem[]{brown03} Brown, T.~M., Ferguson, H.~C., Smith, E., Bowers, C.~W., Kimble, R.~A., Renzini, A., \& Rich, R.~M., 2003, ApJ, 584, L69
\bibitem[]{bruzual03} Bruzual, G., \& Charlot, S., 2003, MNRAS, 344, 1000
\bibitem[Busso et al.~ 2005]{busso05} Busso, G., Moehler, S., Zoccali, M., Heber, U., \& Yi, S. K., 2005, ApJ, 633, L29
\bibitem[]{kuijken01} Kuijken, K., \& Rich, R.~M., 2001, AAS, 199, 9113
\bibitem[]{lejeune97} Lejeune, Th., Cuisinier, F., \& Buser, R., 1997, A\&AS, 125, 229 
\bibitem[]{peterson01} Peterson, R.~C., Terndrup, D.~M., Sadler, E.~M., \& Walker, A.~R., 2001 ApJ, 547, 240
\bibitem[]{pickles98} Pickles, A.~J., 1998, PASP, 110, 863
\bibitem[]{rich93} Rich, R.~M., Minniti, D., \& Liebert, J.~W, 1993, ASPC, 50, 231
\bibitem[]{rich05} Rich, R.~M., Salim, S., Brinchmann, J., Charlot, S., and 24 coauthors, 2005, ApJ, 619, L107 
\bibitem[]{rich05b} Rich, R.~M., \& Origlia, L., 2005, ApJ, 634, 1293
\bibitem[Santos et al.~1995]{santos95} Santos, J.~F.~C.~Jr., Bica, E., Dottori, H., Ortolani, S., \& Barbuy, B., 1995, A\&A, 303, 753
\bibitem[]{schiavon05} Schiavon, R.~P., Rose, J.~A., Courteau, S., \& MacArthur, L.~A., 2005, ApJS, 160, 163
\bibitem[Schlegel et al.~1998]{schlegel98} Schlegel, D.~J., Finkbeiner, \& D.~P., Davis, M., 1998, ApJ, 500, 525
\bibitem[Vallenari et al.~2000]{vallenari00} Vallenari, A., Bertelli, G., \& Schmidtobreick, L., 2000, A\&A, 361, 73
\bibitem[Vallenari et al.~2006]{vallenari06} Vallenari, A., Pasetto, S., Bertelli, G., Chiosi, C., Spagna, A., \& Lattanzi, M., 2006, A\&A, 451, 125
\bibitem[Zoccali et al.~2003]{zoccali03} Zoccali, M., Renzini, A., Ortolani, S., Greggio, L., Saviane, I., Cassisi, S., Rejkuba, M., Barbuy, B., Rich, R. M., \& Bica, E., 2003, A\&A, 399, 931
\bibitem[]{zoccali06} Zoccali, M., Lecureur, A., Barbuy, B., Hill, V., Renzini, A., Minniti, D., Momany, Y., Gomez, A., \& Ortolani, S., 2006, A\&A, 457, L1

\end{thebibliography}
\end{document}